\documentstyle[preprint,prb,aps]{revtex}

\def\AmS{{\protect\the\textfont2
        A\kern-.1667em\lower.5ex\hbox{M}\kern-.125emS}}

\makeatletter
\def\thepage{1-\@arabic\c@page}
\def\@pnumwidth{2em}
\makeatother
\begin{document}
\draft

\title{NMR Study of MnSi under Pressure}

\author{C.~Thessieu$^*$, K.~Ishida$^*$, Y.~Kitaoka$^*$, K.~Asayama$^*$,
G.~Lapertot$^{\dag}$}
\address{$^*$ Department of Material Physics, Faculty of Engineering Science,\\
Osaka University, Toyonaka, Osaka 560}
\address{$^{\ddag}$ Centre d'\'{E}tudes Nucl\'{e}aires de Grenoble, SPSMS, Grenoble, France}
\date{August 25, 1997}
\maketitle

\begin{abstract}
The magnetic phase diagram $T_{\text{c}}(P)$ of the weak itinerant
helimagnetic compound MnSi is reviewed upon a Nuclear Magnetic Resonance
(NMR) experiment. We present a systematic study on the evolution of the NMR
spin echo signal at $T=1.4$~K of the non-magnetic silicone sites, $^{29}$Si,
up to 17.8~kbar. The pressure effect is interpreted as a weak variation of
the local electronic spin polarization and the coexistence of magnetic and
non magnetic Mn atoms under pressure. From the volume dependence of $f_0 $%
($P$), we show notably that a local magnetic order remains above the critical
pressure of $P_{\text{c}}=14.8$~kbar.
\end{abstract}


\section{Introduction}

A lot of studies have shown that in the strongly correlated
electronic systems (the heavy fermions systems with the 4f and 5f electrons
as well as the transition metals with the wider energy band of the 3d
electrons), the ground state is very sensitive to the inter-atomic distance
between the magnetic atoms. Indeed the mechanisms are different depending on
the system, the tune of the unit cell lattice parameters, the electronic
bandwidth and the magnetic exchange energy by an external parameter $\delta$
(which can represent the chemical substitution, $x$, the applied pressure, $P$,
or the applied magnetic field, $H$) give rises to a large variety of ground
states (paramagnetism, magnetic order, superconductivity or spin glass
state,...). The magnetic phase diagram, $T_{\text{c}}(P)$, of the
helimagnetic spin-polarised compound MnSi is well-known since a couple of
years now and several authors have elucidated main of its macroscopic
features \cite{PFL94,PFL95,THE95,THE97}.
Under pressure the Curie temperature decreases monotonously
from 30~K at ambient pressure down to 0 at the critical pressure of
$P_{\text{c}}=14.8$~kbar inducing a quantum critical phase transition (QCPT)
from a spin polarised state to a paramagnetic state. However it remains at
least two questions concerning the evolution of the magnetic properties of
this compound under pressure. The first one concerns the evolution of the
magnetization when the system approaches the quantum critical point. The
second is related to the exact nature of the ground state just above
$P_{\text{c}}$. We address the problem of the long range magnetic order
disappearance under pressure from a microscopic standpoint in the case of a
pure magnetic metal (i.e. without disorder effect induces by the chemical
substitution). In this study we bring answers to these questions
illustrating that the Nuclear Magnetic Resonance is a powerful tool to deal
with the magnetic instability under pressure.

\section{Experimental Details}

For this study a single crystal has been crushed in fine
powder (grain sizes less than 20~$\mu$m) in order to ensure the penetration
of the radiofrequency pulsed field. Our sample has a residual resistivity
ratio of 40 and a residual resistivity at low temperature of roughly 4~$\mu
\Omega .cm$. Preliminarily, we have verified from electrical resistivity
measurements (AC 4 terminals method) that the bulk sample exhibits the same
magnetic features under pressure (e.g. variation of $T_{\text{c}}(P)$ and
the disappearance of the magnetic order at 14.8~kbar) than the one reported
in the literature \cite{PFL94,PFL95,THE95}. The pressure was generated by a ''classical''
clamp-type pressure cell composed of a Be-Cu outer and inner parts. The
pressure transmitting medium was an equal proportion of fluorinert (type 77
and 70) which becomes solid at temperatures below 200~K but remains liquid at
room temperature and in the pressure range of interest. Every cooling
procedure was made slowly enough to avoid stresses in the transmitting
medium and thus inhomogeneous pressure distribution around the sample. The
pressure calibration at low temperature was based on the pressure shift of
the superconducting transition temperature of a bulk tin sample measured by
a 4 probes AC electrical resistivity measurement. The pressure has been
changed in a very systematic way from ambient pressure to the highest
pressure reached during this experiment (17.8~kbar) and we estimate that the
accuracy in the relative pressure change is equal to $\pm $0.1~kbar. The NMR
experiment was performed under zero external field and at 1.4~K using a
phase-coherent pulsed spectrometer. The NMR spectrum was obtained by
plotting the spin-echo intensity as function of frequency with an 20~kHz
interval. For each frequency the NMR conditions (mainly the pulses width
whereas the radiofrequency field amplitude was kept constant) have been
adjusted to maximize the spin echo amplitude.

\section{Results and Analysis}

Previous NMR studies at ambient pressure have reported that in
MnSi the $^{29}$Si NMR signal was observed around $f_0$=19.880~MHz in the
helical magnetic state and at $T$=1.4~K\cite{MOT78}. Below $T_{\text{c}}$,
the magnetic moments at the Mn sites induces a polarization of the
conduction electron (electrons of the s or p shells) at the Si sites. As a
result the transferred hyperfine field acts on the Si nuclei through the
Fermi contact term or the spin-dipole interaction. From this experiment, the
transferred hyperfine field was estimated to be 2.35~Tesla (ambient
pressure) with the gyromagnetic ratio $^{29}\gamma$=8.4578~MHz/Tesla.
Moreover the hyperfine coupling constant was calculated such as
$A_{\text{hf}}$=2.35 Tesla/0.389 $\mu _{\text{B}}$=60.42~kOe/$\mu _{\text{B}}$,
which is compatible with the value obtained from the $\kappa-\chi$ plot in
the paramagnetic state. This agreement assures that the Si NMR can probe the
microscopic magnetic properties inherent to MnSi. By contrast, the Mn NMR
spectrum below $T_{\text{c}}$ exhibits a very broad feature due to the
combined effect of the helical magnetic structure and to a very large
anisotropic hyperfine interaction\cite{MOT78}. Thus the analysis of the
magnetic properties under pressure from this nuclei standpoint is not an
easy task. The Fig.\ \ref{fig1} represents the pressure variationof $^{29}f_0$ at
T=1.4~K, from ambient pressure up to 16.1~kbar. $^{29}f_0$ decreases from
19.880~MHz at ambient pressure to 14.500~MHz for 16.1~kbar pressure above
which the NMR signal disappears (for a pressure of 16.3~kbar no signal was
observed). With increasing pressure, the full width at the half maximum
(FWHM) increases from 200~kHz up to 1.4~MHz. It is noted that the general
spectrum's shape is not modified under pressure conserving a Gaussian-like
shape.

In table\ \ref{table1} are summarized the pressure dependence of
$T_{\text{c}}$, magnetization $M _{\text{s}}$ and $f_0$. The Curie
temperature, $T_{\text{c}}$, is extracted from the anomaly observed in the
$M$-$T$ curves (not shown in this paper), the saturated magnetization,
$M_{\text{s}}$, from the Arrot plot. The hyperfine coupling constant,
$A_{\text{hf}}$, is calculated such as
$A_{\text{hf}}^{\text{Si}}=f_0$/($\gamma \times M_{\text{s}}$). There is a large contrast between the
pressure effect on the Curie temperature and on the saturated magnetization
or the frequency resonance and we underline that the theoretical variation
predicted by the renormalized spin fluctuation theory\cite{MOR85}
$T_{\text{c}} \propto M^{3/2}_{\text{s}}$
is not well respected
under pressure in the case of MnSi. The decrease in $M_{\text{s}}$ is
scaled with the one of $f_0$. As a matter of fact, $M_{\text{s}}$ vs
$^{29}f_0$ plotted with the pressure as an implicit parameter is fitted linearly
and the hyperfine coupling constant is deduced to be volume independant.

Surprisingly we have observed the persistence of the spin echo
signal even above $P_{\text{c}}$=14.8~kbar identified as the critical
pressure at which the long range magnetic disappears from the bulk
measurements\cite{PFL94}. The fact that the Si NMR signal was observable at
zero-field assures the persistence of a local magnetic order over $P_{%
\text{c}}$ from a microscopic point of view.

To estimate the spectrum intensity, the spin echo intensity
was fitted by $I(\tau)=I(0)\exp(-2\tau/T_2)$ where $\tau$ is the time
duration between the two pulses and $T_2$ the spin echo decay time and $I(0)$
is reported on the Fig.\ \ref{fig2} as function of the frequency. The area of this
spectrum has been calculated by fitting this latter with the superposition
of two Gaussian functions. As a result it turns out that this area is
proportional to the total fraction of Si nuclei included in magnetic
domains. The integrated intensity of the NMR spectrum at 1.4~K decreases
linearly in a range of 7.5 - 12.5~kbar followed by a weak pressure
dependence up to 16.1~kbar. Unexpectedly exceeding 7.5~kbar, non-magnetic
domains start to be induced by pressure. This results in a decrease of the
NMR spectrum intensity. That is to say that the helicoidally odered state is
partially suppressed over 7.5~kbar to presummably change into paramagnetic.
However above $P_{\text{c}}$, local helimagnetic domains survive which
disappears abruptaly at 16.1~kbar. For this pressure, the domain size
is anticipated to become smaller than the spatial periodicity of the helical
structure ($\lambda=2\pi/Q$) leading to a shrinkage of domains.

\section{Conclusion}

We have carried out a $^{29}$Si NMR experiment at zero-field
in order to study the helicoidally ordered to paramagnetic quantum phase
transition induced by applying pressure up to 17.8~kbar. The $^{29}$Si NMR
frequency has revealed a very weak variation of the transferred hyperfine
field and hence of the local magnetization on the Mn sites. Moreover the
persistence of a $^{29}$Si NMR signal over the critical pressure of
14.8~kbar has proved that the magnetic order is not destroyed from a local
point of view even though the phase transition is not identified
macroscopically. In this note we do not deal with the physical origin of the
local magnetic order persisting above 14.8~kbar. Recently Buzdin and
Meurdesoif \cite{BUZ97} have demonstrated theoretically that in the peculiar
case of a helimagnetic modulated phase the presence of local defects
(impurities, dislocations and defects) play a keyrole in the formation of
unusual local magnetic order centered on these defects. At this stage, it is
worth wondering if the persistence of a local magnetic order above $P_{%
\text{c}}$ is not the mark of the local defects. In the peculiar case of
MnSi the influence of these local defects may play an important role.

\begin{figure}
\caption{The main figure compares the
pressure effect on the Curie temperature $T_{\rm c}$(---) and the frequency
resonance $f _0$ (($\bullet $) increasing $P$ and ($\circ $) releasing $P$). The
inset presents the pressure effect on the resonance line of the $^{29}$Si
nuclei at $T$=1.4~K (Ambient, 3.65, 5.77, 7.57, 10.29, 12.99 and 15.64~kbar).}
\label{fig1}
\end{figure}

\begin{figure}
\caption{ Evolution of the resonance line
intensity with the applied pressure. The plain line represents schematically
the magnetic phase diagram of MnSi.}
\label{fig2}
\end{figure}

\begin{table}
\caption{Pressure dependence of macroscopic and microscopic quantities in
MnSi.}
\label{table1}
\begin{tabular}{dcccc}
$P$[kbar] & $T_{\text{c}}$[K] &
$M_{\text{s}}$[$\mu_{\text{B}}$/Mn] &
$f_0$[MHz] & $A^{^{29}\text{Si}}_{\text{hf}}$[kOe/$\mu_{\text{B}}$] \\ \tableline
Ambient & 29.07 & 0.389 & 19.880 & 60.42 \\ 
3.6 & 24.01 & 0.374 & 19.270 & 60.91 \\ 
6.6 & 19.60 & 0.359 & 18.504 & 60.94 \\ 
8.7 & 16.27 & 0.348 & 17.840 & 60.61 \\ 
10.2 & 13.50 & 0.342 & 17.340 & 59.94 \\ 
\end{tabular}
\end{table}

\end{document}